\documentclass
[preprint,prl,10pt,twocolumn,lengthcheck,bibnotes,noshowpacs]{revtex4}%
\usepackage{amsfonts}
\usepackage{amsmath}
\usepackage{amssymb}
\usepackage{graphicx}%
\begin{document}
\preprint{UATP/07-02}
\title{The Concept of Entropy and its Concavity for a Finite Protein in its
Environment:\ An exact study on a Square Lattice }
\author{P.D. Gujrati, and Bradley Lambeth, Jr. }
\email{pdg@arjun.physics.uakron.edu}
\affiliation{The Departmentof Physics, The Department of Polymer Science, The University of
Akron, Akron, OH, 44325}
\date{\today}

\pacs{PACS number}

\begin{abstract}
We consider a general lattice model of a finite protein in its environment and
calculate its Boltzmann entropy $S(E)$ as a function of its energy $E$ in a
microcanonical ensemble, and Gibbs entropy $\overline{S}(\overline{E})$ as a
function of its average energy $\overline{E}$ in a canonical ensemble by exact
enumeration on a square lattice. We find that because of the finite size of
the protein, (i) the two are very different and $\overline{S}(\overline
{E})>S(\overline{E})$, (ii) $S(E)$ need not be concave while $\overline
{S}(\overline{E})$ is, and (iii) $\overline{S}(\overline{E})$ is relevant for
experiments but not $S(E)$, even though $S(E)$ is conceptually more useful. We
discuss the consequences of these differences. The results are general and
applicable to all finite systems.

\end{abstract}
\maketitle

Self-assembling small proteins are a prime example of small systems, and can
fold into their native states (of minimum free energy) without any chaperones.
They have been extensively investigated recently using lattice models by
thermodynamic principles \cite{Anfinsen}. The smallest known natural protein
is Trp-Cage derived from the saliva of Gila monsters and has only 20 residues.
Their first-principle study requires short ranged model energetics that, while
remaining independent of the thermodynamic state of the protein such as its
conformation, temperature $T,$ pressure $P$, etc., determine the native
state(s), and has to be judiciously chosen to give a unique and right native
state. It should be stressed that proteins in Nature are never isolated but
always occur in an environment such as a cell controlled by the temperature
$T$. Thus, the proper way to study proteins is to consider the canonical
ensemble (CE), and not the microcanonical ensemble\ (ME). Moreover, the two
ensembles are most probably not \emph{equivalent} for a finite system. Despite
this, investigations using ME are very common for proteins. Therefore, their
predictions must be carefully examined and compared with those from CE,
keeping in mind their possible non-equivalence. Unfortunately, this does not
seem to be practiced in the field, which as we will establish here may be
quite dangerous for finite systems such as proteins.

The ME entropy is given by the Boltzmann relation $S(E)\equiv\ln W(E),$\ where
$W(E)$ is the number of protein conformations of energy $E.$ Since folding is
a conformational change into the native state, the conformational entropy
$S(E)$ is believed to play a central role in determining the way folding
occurs into compact native states along a very large number of microscopic
pathways that connect them to myriad unfolded conformations. It also
characterizes the potential energy landscape \cite{Miller,Wales,Sali}. It is a
well-established tenant of \emph{macroscopic} thermodynamics that
$W(E)$\ decreases with falling energy $E$ as folding proceeds ($\partial
S/\partial E\geq0$); consequently, the energy landscape is expected to possess
a structure that narrows down with falling energy, such as a funnel
\cite{Guj0412548}. It is known that the entire thermodynamics is contained in
$S(E)$, which must be \emph{concave} \cite{note1} for a macroscopic system.
This concavity is built-in in the random energy model \cite{Derrida}, which
has been extensively employed for proteins; see \cite{Sali} for example, which
also shows that the energy gap above the ground state (lowest energy state) is
crucial for foldability. The resulting lack of concavity has been used as a
sign of a first-order folding transition for small proteins by several
workers. It should be noted that there are other idealized physical models
such as the KDP model that freeze into the ground state at finite $T$ due to a
similar gridlock \cite{KDP,Nagle}.

A proper model of a\ protein should satisfy certain principles
\cite{Scheraga0}, one of which is the requirement of cooperativity. The
sequence of residues also plays an important role in determining the native
state \cite{Shakhnovich}. However, there is no consensus for general
energetics to describe all proteins, and there remains a certain amount of
freedom in the choice, at least in modeling. It is widely recognized that
secondary structures are also important in the folding process \cite{Dill}.
The simplest model is the standard model, which classifies the 20 different
residues into two, H (hydrophobic) and P (hydrophilic), and allows only
nearest-neighbor attractive HH interaction $e_{\text{HH}}$ (set $=-1$ in some
predetermined unit) to provide good hydrophobic cores \cite{Dill}; however,
consideration of local energetics of the 20 kinds of residues \cite{Miyazawa}
is also common. It is found that the introduction of multi-body interactions
enhances cooperativity \cite{Kolinski}, and should not be neglected. It is
important, therefore, to investigate the energetics effects on the form of
$S(E)$, which to the best of our knowledge has not been studied carefully.

The direct experimental approaches (primarily, X-ray crystallography or NMR
spectroscopy) to determine energetics requires information about the
\emph{typical} conformation associated with the average energy $\overline
{E}(T)$. Thus, CE must be used to determine the dependence of the
\emph{canonical} entropy $S(T)$ on $\overline{E}$, given by the Gibbsian
relation $S(T)=-\sum p(\Gamma,T)\ln p(\Gamma,T),$ where $p(\Gamma,T)$\ is the
probability to be in the conformation $\Gamma$ at $T.$ For a macroscopic
system, $S(\overline{E})$ and $S(T)$ are the same so that $S(T)$ allows us to
identify conformations of average energy $\overline{E}.$ Their equality is
crucial for the direct experimental approaches in which conformations
associated with $\overline{E}$ need to be identified as \emph{typical}. Thus,
it is also important to verify if the two entropies are the same for finite
proteins that are of interest here. If not true, the interpretation of
experimental data for the energetics could be \emph{incorrect}. This will
become a limitation of the direct experimental techniques.

\textbf{Model }The interplay of intra-protein molecular interactions, the
interaction with the surrounding, and the residue sequence to give rise to the
folded native state is quite intricate and far from a basic understanding at
present; much remains to be understood. It has been argued that conflicts
among interactions also play a significant role in folding \cite{Clementi}. In
general, the model should contain various interactions relevant not only for
protein folding and various secondary substructures like helix formation in
the native state, but also for proteins considered as semi-flexible
heteropolymers \cite{GujSemiflex} with certain specific sequences
\cite{Venkat}. The model should also contain solvation effect, as all protein
activity occurs in the presence of water or solvent. In this work, we use such
a model, which has been investigated by us recently \cite{PDGProtein} in
different limits one of which is the standard model described above. Here, we
only report some unexpected results for finite proteins, which have not been
noted earlier to the best of our knowledge. It should, however, be recognized
that finite proteins cannot undergo a sharp folding transition. This issue is
not relevant as we are only interested in comparing $S(E)$ and $\overline
{S}(\overline{E}).$%
\begin{figure}
[ptb]
\begin{center}
\includegraphics[
trim=1.560227in 2.789137in 2.043368in 3.513824in,
height=2.6308in,
width=2.7544in
]%
{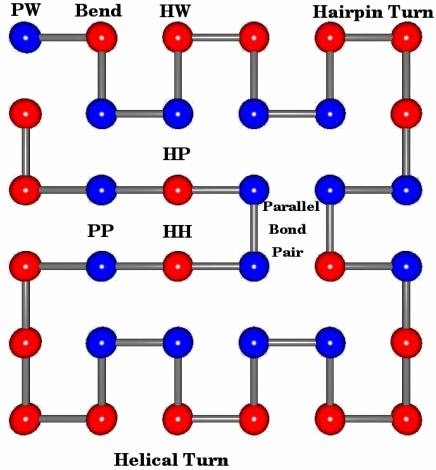}%
\caption{A 2-d model of a finite protein on a square lattice. The red spheres
represent hydrophobic sites and the blue spheres represent hydrophilic sites.}%
\label{F1}%
\end{center}
\end{figure}
\qquad

We consider a protein with $M$ residues in a given sequence on a square
lattice, with one of its ends fixed at the origin so that the total number of
conformations $W$ for a finite protein remains finite even on an infinite
lattice. We generalize a recent model \cite{GujSemiflex}, in which the number
of bends $N_{\text{b}},$ pairs of parallel bonds $N_{\text{p}},$ and hairpin
turns $N_{\text{hp}}$\ characterize the semiflexibility; see Fig. \ref{F1},
where we show a protein in its compact form so that all the solvent molecules
(W) such as water are expelled from the inside and surround the protein. We do
not allow any free volume. The red spheres denote hydrophobic residues (H) and
blue spheres denote hydrophilic (i.e., polar) residues (P). The
nearest-neighbor distinct pairs PP, HH, HP, PW and HW between the residues and
the water are also shown, but not the contact WW. Only three out of these six
contacts are independent on the lattice \cite{Guj2003}, which we take to be
HH, HW, and HP pairs. A bend is where the protein deviates from its collinear
path. Each hairpin turn requires two consecutive bends in the same (clockwise
or counterclockwise) direction; see Fig.\ref{F1}. Two parallel bonds one
lattice spacing apart form a pair (p). We also consider the number of helical
turns $N_{\text{hl}}.$ On a square lattice, a "helical turn" is interpreted as
two consecutive hairpin turns in opposite directions; see Fig.\ref{F1}. The
corresponding energies are $e_{\text{b}},$ $e_{\text{P}},$ $e_{\text{hp}}$,
and $e_{\text{hl}},$ respectively$.$ The pair interaction energies are
$e_{\text{HH}}=-1,$\ $e_{\text{HW}},$ and $e_{\text{HP}},$\ and the pair
numbers are $N_{\text{HH}},$ $N_{\text{HW}},$ and $N_{\text{HP}},$
corresponding to the HH, HW, and HP, respectively. respectively. We let
$\mathbf{e}$ denote the entire ordered set $\{e_{i}\},$with $i$ ordered as
b,p,hp,hl,HH,HW, and HP, and $\mathbf{e}^{\prime}$ the ordered set $\left\{
e_{i}\right\}  $ excluding $e_{\text{HH}}(=-1).$ Similarly, $\mathbf{N}%
\equiv\mathbf{N}(\Gamma)\equiv\{N_{i}(\Gamma)\},$ and $\mathbf{N}^{\prime}$
denotes all $\left\{  N_{i}\right\}  $ but $N_{\text{HH}}.$ The three most
often energy choices we have made are:\ (A) $\mathbf{e}^{\prime}=0,$ (B)
$\mathbf{e}^{\prime}=(a,-a,-2a,-a,25a,5a),$ $a=1/50(<<1),$ (C)$~\mathbf{e}%
^{\prime}=(\ b,-b,-b,-b,2b,b),b=1/3(\simeq1).$ The standard model is (A). In
the model (B), we have most other interactions much weaker than $\left\vert
e_{\text{HH}}\right\vert $, while they are comparable to $\left\vert
e_{\text{HH}}\right\vert $ in the model (C). Thus, (B) is closer to (A) than
(C) is. Despite this, we will see that (B) and (C) behave very different from
(A). It should be noted that $W$ does not depend on the model; it is its
partition into $W(E)$ that depends on the model. Thus, the shape of the energy
landscape changes from model to model, but not its total "area" which is given
by $W$ \cite{Guj0412548}. We have also considered random, ordered and fixed
sequences. We consider \emph{compact} and \emph{unconstrained} protein
conformations \cite{PDGProtein} separately. We have found that in the majority
of cases that we have investigated, the sequence containing a repetition of
PPHH gives rise to the lowest energy or very close to it.

The energy of a given conformation $\Gamma$ is
\begin{equation}
E(\Gamma)\mathcal{\equiv}\mathbf{e}\cdot\mathbf{N}(\Gamma)\equiv%
{\textstyle\sum}
e_{i}N_{i}(\Gamma). \label{Energy}%
\end{equation}
We partition $W$ according to $\mathbf{N}$ or $E$, so that $W\equiv
\sum_{\mathbf{N}}W(\mathbf{N})\equiv\sum_{E}W(E)$, where $W(\mathbf{N})$ [or
$W(E)$] is the number of conformations for a given set $\mathbf{N}$\ [or $E$].
On a lattice, $E$ remains a discrete variable, but this fact is not important
for our final conclusions as we will discuss below. In the standard model,
$E=-N_{\text{HH}}$. It is clear from%
\begin{equation}
W(N_{\text{HH}})\equiv%
{\textstyle\sum}
W(N_{\text{HH}},\mathbf{N}^{\prime}), \label{W-relation}%
\end{equation}
that the entropy $S(N_{\text{HH}})=\ln W(N_{\text{HH}})$ for a given
$N_{\text{HH}},$ regardless of $\mathbf{N}^{\prime}$,\ is maximum in the
standard model \cite{Gujrati1} and provides a possible justification of the
observation made in \cite{Kolinski}. A protein with a given $N_{\text{HH}}$
will probe many more states in the standard model, where there is no energetic
penalty to explore all possible $\mathbf{N}^{\prime}$,\ than in any other
model with energetic penalty, which then slows down its approach to the native
state. Thus, it is important to have non-zero $\mathbf{e}^{\prime}$ to step up
the approach to the native state. (It is highly likely that the native states
in different models are different, but this does not affect the above
conclusion.) \ There is another important consequence of $\mathbf{e}^{\prime
}=0.$ The fluctuations in the corresponding $N_{i}$ are maximum as there is no
penalty no matter what $\mathbf{N}^{\prime}$ is. The protein will spend a lot
of time probing a large number of conformations corresponding to the maximum
fluctuations in $\mathbf{N}^{\prime}.$\ This also suggests that we need to go
beyond the standard model to describe proteins that fold fast.

The canonical probability distribution for $\Gamma$ is $p(\Gamma,T)\equiv
e^{-\beta E(\Gamma)}/Z(T),$ where
\begin{equation}
Z(T)\equiv%
{\textstyle\sum\limits_{\Gamma}}
e^{-\beta E(\Gamma)}\equiv%
{\textstyle\sum\limits_{E}}
W(E)e^{-\beta E},\label{PF}%
\end{equation}
the partition function, describes the finite protein thermodynamics; here,
$\beta$ $\equiv1/T$ (we set the Boltzmann constant $k_{\text{B}}=1).$ The
distribution $p(\Gamma)$ can be used to define the average $<>$ of any
thermodynamic quantity (also denoted by an overbar in the following) such as
$\overline{\mathbf{N}}(T)\equiv<\mathbf{N>},$ and $\overline{E}(T)\equiv
<E>\equiv\mathbf{e}\cdot\overline{\mathbf{N}}(T)$ $[\overline{e}%
\equiv\overline{E}/M]$. The free energy $F(T)\equiv-T\ln Z(T)\ $gives the
canonical entropy $S(T)\equiv-\partial F(T)/\partial T$ $,$ which satisfies
the conventional thermodynamic relation $F(T)\equiv\overline{E}(T)-TS(T),$ and
the Gibbsian relation quoted above, as can be easily checked$.$ Both $S(T)$
and $\overline{E}(T)$\ are \emph{continuous function} (except possibly at a
phase transition, which is not relevant here as we are dealing with a finite
protein) of the continuous variable $T.$ Moreover, $F(T)$ is monotonically
decreasing with $T$ as expected.

\begin{figure}
[ptb]
\begin{center}
\includegraphics[
trim=0.838367in 3.084971in 1.636813in 3.512759in,
height=2.2122in,
width=3.0908in
]%
{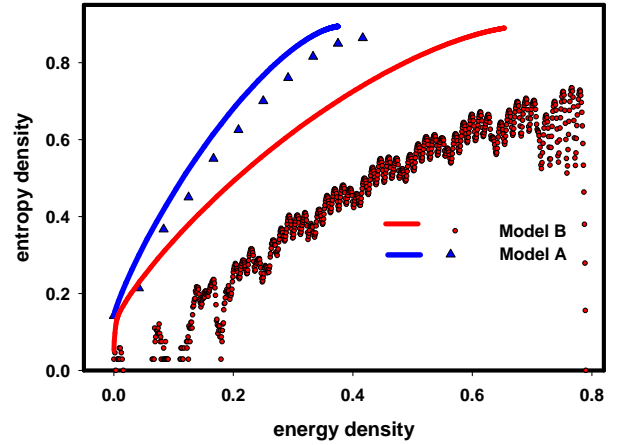}%
\caption{Continuous $\overline{s}(\overline{e})$ (blue and red curves), and
discrete $s(e)$ (blue and red points) for a given sequence $(M=24$,
unrestricted$);$. The bands in $s$ become more pronounced and their
separations decrease as $M$ increases. Note a clear band in $s$ at low
energies and the native state, disjoint from the rest of the bands. }%
\label{F2}%
\end{center}
\end{figure}

Since the derivative $\partial\overline{E}/\partial T$ is non-negative as can
be easily checked, $\overline{E}$ can be \emph{inverted} to express $T$ as a
function $T(\overline{e}),$ which then allows us to express $S(T)$ as an
explicit function $\overline{S}(\overline{E})\equiv S[T(\overline{e})]$\ of
$\overline{E}.$ The entropy $\overline{S}(\overline{E})$ can be thought of as
the \emph{canonical equivalence} of the microcanonical entropy $S(E).$
However, they are two \emph{different} quantities for finite proteins. In the
first place, $S(E)$ is a discrete function since $E\ $is discrete, while
$\overline{S}(\overline{E})$ is a continuous function since $\overline{E}$ is
continuous. In the second place,$\ \overline{S}(\overline{E})\geq
S(\overline{E}),$ the equality holding as $M\rightarrow\infty$
\cite{Guj0412548}. To demonstrate this, let us assume that $E=\overline{E}$ is
one of the energies in the sum in (\ref{PF}). We then rewrite \ $\overline
{S}(\overline{E})=\ln Z+\overline{E}/T,$ and evaluate $\overline{W}%
(\overline{E})\equiv\exp[\overline{S}(\overline{E})]$: \
\begin{equation}
\overline{W}(\overline{E})=W(\overline{E})+%
{\textstyle\sum\limits_{E\neq\overline{E}}}
W(E)e^{-\beta(E-\overline{E})};\label{ContW}%
\end{equation}
hence, $\exp[\overline{S}(\overline{E})]\geq\exp[W(\overline{E})]$ as asserted
above.\ The difference between them is due to the non-negative last term in
(\ref{ContW}), which vanishes as $N\rightarrow\infty.$ In case, $\overline{E}$
is not one of the energies in the sum, we can use a suitable interpolation to
define $\overline{W}(\overline{E}),$ without affecting the conclusion
\cite{PDGProtein}. The above proof does not depend on the discrete nature of
the energies in ME; thus, it is also valid for continuum models. We show in
Fig.\ref{F2} the exactly enumerated entropies per residue $s(e)\equiv
(1/M)S(E)$ (red curve as a guide through discrete points) and $\overline
{s}(\overline{e})\equiv(1/M)\overline{S}(\overline{E})$ (blue curve) for the
model (B) ($M=24;$ unrestricted conformations) as a function of the discrete
variable $e\equiv E/M$ or $\overline{e}.$ In addition, we also see a distinct
\emph{band structure} in $s(e)$ that gives rise to regions of non-concavity
\cite{note1}, which is related to the nature of the interactions and has no
implication for any phase transition as we will discuss below.

It is easily seen that the canonical entropy function satisfies the
conventional thermodynamic relation \cite{Guj0412548}
\begin{equation}
\partial\overline{S}(\overline{E})/\partial\overline{E}=1/T,\label{T-relation}%
\end{equation}
and is, therefore, concave ($\partial^{2}\overline{S}(\overline{E}%
)/\partial\overline{E}^{2}<0$) \cite{note1}$.$ On the other hand, the
microcanonical entropy need \emph{not }be concave; see Fig.\ref{F2}, where the
bands seen in $s(e)$ have both positive and negative slopes, which is in
contradiction with (\ref{T-relation}) valid for $\overline{s}(\overline{e}).$
The non-concave $S(E)$ does not violate finite system thermodynamics. The
canonical entropy is the physical entropy for proteins in its environment and
remains concave in Fig. \ref{F2} as required by thermodynamic stability.%

To understand the absence of concavity, we first consider the model (A). In
all cases that we have studied \cite{PDGProtein}, $S(E)=S(N_{\text{HH}})$ is
found to be concave. The number of states $W(N_{\text{HH}})$ can be
partitioned into $W(N_{\text{HH}},\mathbf{N}^{\prime})$; see (\ref{W-relation}%
). In the model (B), $\mathbf{e}^{\prime}\simeq0;$ therefore, most of the
conformations in\ $W(N_{\text{HH}})$ have energies that are close to
$-N_{\text{HH}};$ some of them will have energies that are outside the range
($-N_{\text{HH}}-1,-N_{\text{HH}}+1).$ The resulting $S(E)$ associated with
this $N_{\text{HH}}$ is almost concave, as seen in each band in Fig.\ref{F2}.
This then gives rise to the lack of concavity in the region where two nearby
bands overlap. The number of bands equals the number of possible values of
$N_{\text{HH}}$ in the model (A). These convex portions of $s(e)$ disappear
and $s(e)$ approach $\overline{s}(\overline{e})$ from below as $M\rightarrow
\infty$ $\ $\cite{Guj0412548}. But for finite systems, the convex regions
persists. The band structure persists for all sequences that we have checked.
The additional energies in the model (C) provide enough spread for bands to
overlap; this reduces the size of convex regions. Even here, we have found
that the band nature survives at the upper and the lower ends of the energy.
Thus, we are confident that convex regions in $S(E)$ will exist in any
realistic model of a protein. Their presence, however, does not imply any
phase transition, as $\overline{S}(\overline{E})$ is always concave. This is
true even though we note from Fig.\ref{F2}, that there is a clear gap at the
lowest energy. The energy gap causes convexity in $S(E),$ but not in
$\overline{S}(\overline{E}).$

Because of conformational changes during folding, the folding is believed to
be governed by the multiplicity $W(E),$ which in turn governs the energy
landscape for which $W(E)$ represents the "surface area" of the hypersurface
of the landscape at energy $E$ \cite{Guj0412548}:\ each point on the
hypersurface represents a conformation. The lack of concavity discovered here
has a profound effect on the \emph{shape} of the landscape. It no longer
narrows down as $E$ decreases. It will be interesting to pursue the
consequences of this shape modification. This is beyond the scope of the
present work, but we hope to consider it elsewhere. It is evident, and as
discussed above, several different $\mathbf{N}$ will usually mix together for
a given $E$, except in the model (A) in which $E=-N_{\text{HH}}$ so that
$W(E)=W(N_{\text{HH}}).$ There will be a certain landscape topology for the
standard model, which will change with $\mathbf{e}^{\prime}$. From
(\ref{W-relation}), it is evident that the landscape will become drastically
narrower for $\mathbf{e}^{\prime}\neq0.$ The total "surface area" $W$ of the
landscape does not change with $\mathbf{e}^{\prime}$, even though the allowed
energies change and they become closer$.$ The landscape narrowing and
closeness of energies at constant $W$ make the approach to native
state\ presumably $\overline{E}$more directional and fast.

Since it is CE that is relevant for a real protein in its environment, it is
the canonical multiplicity $\overline{W}(\overline{E})$ that is relevant for
folding. As shown above, it continuously increases with $\overline{E},$ until
we reach at infinite temperatures. Thus, the narrowing of the landscape with
non-zero $\mathbf{e}^{\prime}$ may not be as relevant for protein folding as
the observation that $\overline{W}(\overline{E})>W(\overline{E}).$\ From
(\ref{ContW}), we observe that $\overline{W}(\overline{E})$ gets contribution
from \emph{all} conformations, not just the conformations in $W$
$(\overline{E}).$ Thus, it may be misleading to think that a finite protein at
a given $T$ only probes some typical conformations of average energy
$\overline{E}.$ (It is possible that $\overline{E}$ may not even be an allowed
energy $E$.) It also probes native state, though its probability is going to
be small$.$ As $T$ is reduced, this probability increases. In addition, the
protein restricts its search to effectively a smaller set of conformations,
closer in energy. It would be interesting to follow the consequence(s) of this observation.

In conclusion, we observe that the microcanonical entropy, which dictates the
form of energy landscape, does not satisfy concavity; however, this violation
does not imply any impending phase transition; the latter requires
investigating the behavior of the canonical entropy, which always satisfies
concavity. However, nearly all works on protein thermodynamics have not paid
any attention to this issue. This may be dangerous. The most surprising result
is the tremendous difference between the two entropies: the canonical entropy
is almost twice as big as the microcanonical entropy at intermediate energies,
but much larger at low energies. Its implication for experimental data
interpretation, as noted above, needs to be further pursued.

We would like to thank Andrea Corsi, whose original program code for
enumeration was used and extended for the results presented here. We
gratefully acknowledge the NSF support of this research through the University
of Akron REU Site for Polymer Science (DMR-0352746).


\begin{thebibliography}{99}
\bibitem {note1}The graph of a concave function (discrete or continuous) is
one in which the line connecting its values at any two points must never lie
above the function at intermediate points.
\end{thebibliography}
\end{document}